\documentclass[prb,aps,twocolumn]{revtex4}
\usepackage{epsfig}
\include{psfig}

\begin{document}


\title{\bf Semiclassical theory of current correlations in chaotic
dot-superconductor systems.}  
\author{P. Samuelsson and M. B\"uttiker}
\affiliation{D\'epartement de Physique Th\'eorique, Universit\'e de
Gen\`eve, CH-1211 Gen\`eve 4, Switzerland.}  
\begin{abstract}
We present a semiclassical theory of current correlations in
multiterminal chaotic dot-superconductor junctions, valid in the
absence of the proximity effect in the dot. For a dominating coupling
of the dot to the normal terminals and a nonperfect dot-superconductor
interface, positive cross correlations are found between currents in
the normal terminals. This demonstrates that positive cross
correlations can be described within a semiclassical approach. We show
that the semiclassical approach is equivalent to a quantum mechanical
Green's function approach with suppressed proximity effect in the dot.
\end{abstract}
\maketitle
Over the last decade, there has been an increasing interest in current
correlations in mesoscopic conductors. \cite{Buttikerrew} Compared to
the conductance, the current correlations contain additional
information about the transport properties such as the effective
charge or the statistics of the quasiparticles. Systems such as
e.g. disordered conductors and chaotic quantum dots have been analyzed
extensively, both with quantum mechanical approaches, using random
matrix theory\cite{Beenakker92,vanlangen97} or Green's
functions,\cite{Altshuler94} as well as with semiclassical approaches,
based on the
Boltzman-Langevin\cite{Nagaev92,deJong96,Sukhorukov99,Blanter00,Blanter01}
equation or voltage probe models.\cite{deJong96,vanlangen97,Blanter01}

Recently, current correlations in normal-superconducting systems have
been studied, theoretically \cite{Buttikerrew} as well as
experimentally.\cite{Jehl00} In these systems, the current into the
superconductor is transported, at subgap energies, via Andreev
reflection at the normal-superconducting interface. In the normal
conductor, Andreev reflection induces a proxmity effect which modifies
the transport properties at energies of the order of or below the
Thouless energy.\cite{Lambert} However, at energies well above the
Thouless energy or in the presence of a weak magnetic field in the
normal conductor, the proximity effect is suppressed.

Nagaev and one of the authors\cite{Nagaev01} presented a semiclassical
Boltzman-Langevin approach, in the absence of the proximity effect,
for current correlations in multiterminal diffusive
normal-superconducting junctions, an extension of the corresponding
approach for purely normal systems.\cite{Sukhorukov99} The interfaces
between the normal conductor and the superconductor was assumed to be
perfect. It was shown that the current cross correlations are
manifestly negative, just as in normal mesoscopic
conductors. \cite{Buttiker92} In contrast, it was shown very recently,
taking the proximity effect into account, that in diffusive tunnel
junctions\cite{Boerlin02} and chaotic dot juctions,\cite{Samuelsson02}
the ensemble averaged cross correlations can be positive.

Interestingly, as shown for the chaotic dot juction,
\cite{Samuelsson02} in the limit of strong coupling of the dot to the
normal reservoirs and a nonperfect normal-superconducting interface,
positive cross correlations can survive even in the abscence of the
proximity effect. This suggests that positive current correlations
could be obtained within a semiclassical approach, under the
assumption of a suppressed proximity effect, if nonperfect
normal-superconductor interfaces are assumed.

In this paper we present such a semiclassical theory for multiterminal
chaotic dot-superconductor junctions with arbitrary transparencies of
the contacts between the dot and the normal and superconducting
reservoirs. It confirmes and extends the result of
Ref. [\onlinecite{Samuelsson02}], providing conditions on the contact
widths and transparencies for obtaining positive correlations. We show
that the result is in agreement with the circuit theory of
Refs. [\onlinecite{Nazarov,Boerlin02}] with suppressed proximity
effect in the dot. This provides a simple, semiclassical explanation
for the positive cross correlations.
\begin{figure}[h]
\centerline{\psfig{figure=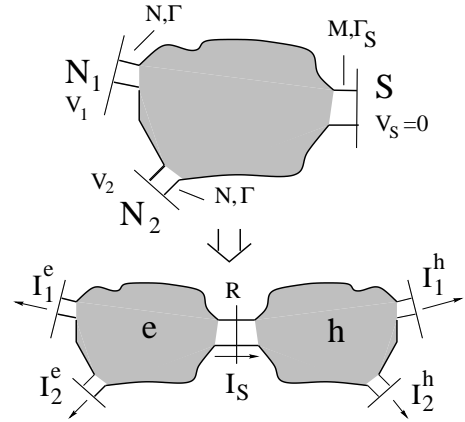,width=6.0cm}}
\caption{Upper figure: A schematic picture of the junction. Lower
figure: Mapping onto an electron-hole junction (see text). The arrows
show the direction of particle current flows.}
\label{fig1}
\end{figure}

A schematic picture of the system is shown in Fig. \ref{fig1}. We
consider a chaotic quantum dot (see Ref. [\onlinecite{Beenakker97}]
for definition) connected to two normal ($N_1$ and $N_2$) and one
superconducting reservoir ($S$) via quantum point contacts.  The
contacts to the normal and superconducting reservoirs have mode
independent transparencies $\Gamma$ and $\Gamma_S$ respectively and
support $N$ and $M$ transverse modes.

The conductances of the point contacts are much larger than the
conductance quanta $2e^2/h$, i.e. $N\Gamma,M\Gamma_S \gg 1$, so
Coulomb blockade effects in the dot can be neglected.  The two normal
reservoirs are held at the potentials $V_1$ and $V_2$ and the
potential of the superconducting reservoir is zero. Inelastic
scattering in the dot is neglected. Throughout the paper, we consider
the case where $V_1,V_2, kT \ll \Delta$, where $\Delta$ is the
superconducting gap. In this case, the probability for Andreev
reflection at the normal-superconductor interface,
$R=\Gamma_S^2/(2-\Gamma_S)^2$, is independent of energy and there is
no single particle transport into the
superconductor. \cite{commentvolt}

The semiclassical approach for current correlations in normal chaotic
dot systems has been developed within a voltage probe approach
\cite{vanlangen97,Blanter01} or equivalently, a minimal correlation
approach. \cite{Blanter00,Blanter01,Oberholzer01} Here we extend it to
a chaotic dot-superconductor system, shown in Fig. \ref{fig1}, by
first considering the {\it particle} current correlations in the
dot-superconductor system. To visualize the particle current flow, it
is helpful to view the chaotic dot-superconductor system as one
``electron dot'' and one ``hole dot'', coupled via Andreev reflection
(see Fig. \ref{fig1}). Due to the abscence of the proximity effect,
the electron and hole dots can be treated as two coupled, independent
dots, and one can calculate the particle current correlations in the
same way as in a purely normal system (Andreev reflection conserves
particle currents). From the particle current correlations, we then
obtain the experimentally relevant {\it charge} current correlations.

We study the zero-frequency correlator
\begin{equation}
P_{jk}=2\int dt~\langle \Delta I_j(t)\Delta I_k(0)\rangle
\label{currcorrexpr}
\end{equation}
between charge currents flowing in the contacts $j=1,2$ to the normal
reservoirs, where $\Delta I_j(t)=I_j(t)-\bar I_j$ is the current
fluctuations in lead $j$ ($\bar I_j$ is the time averaged current). The charge
current $I_j(t)$ is the difference of the electron (e) and hole (h)
particle currents,
\begin{equation}
I_j(t)=I^e_j(t)-I^h_j(t),~~I_j^{\alpha}(t)=\frac{e}{h}\int dE~i_j^{\alpha}
(E,t),
\label{ehcurr}
\end{equation}
where the particle currents densities $i_j^{\alpha}(E,t)=\bar
i_j^{\alpha}(E)+\Delta i_j^{\alpha}(E,t)$. The electron current
density flowing into the superconductor is $i_S^e(E,t) \equiv
i_S(E,t)=-i_S^h(E,t)$.

First, the time average current is studied. Using the requirement of
current conservation at each energy [due to the abscence of inelastic
scattering], the time averaged electron and hole distribution
functions $\bar f_e$ and $\bar f_h$ can be determined. We have for
each dot separately [suppressing the energy notation]
\begin{eqnarray}
\bar i_1^e+\bar i_2^e+\bar i_S&=&0,~\bar i_1^h+\bar
i_2^h-\bar i_S=0.
\label{curr1}
\end{eqnarray}
The currents $\bar i_{j}^{\alpha}$ and $\bar i_S$ flowing through the
point contacts are given by
\begin{eqnarray}
\bar i_{j}^{\alpha}&=&N\Gamma\left[\bar
f^{\alpha}-f^{\alpha}_{j}\right],~ \bar i_S=MR[\bar f^e-\bar f^h],
\label{curr2}
\end{eqnarray}
where $f_j^{\alpha}=[1+\mbox{exp}([E\pm eV_j]/kT)]^{-1}$ is the
distribution function of quasiparticle type $\alpha$ of the normal
reservoir $j$, with $-(+)$ for electrons (holes). Combining
Eq. (\ref{curr1}) with Eq. (\ref{curr2}), we obtain
\begin{eqnarray}
\bar f^e=\frac{MR(f_1^h+f_2^h)+(2N\Gamma+MR)(f_1^e+f_2^e)}{4(N\Gamma+MR)},
\label{distfunc}
\end{eqnarray}
and similarily $\bar f^h=\bar f^{e\rightarrow h}$, where we note that
$\bar f^h(E)=1-\bar f^e(-E)$, as demanded by electron-hole
symmetry. We can then calculate the time averaged currents from
Eq. (\ref{curr2}) and (\ref{ehcurr}). This is however not further
discussed here.

We now turn to the fluctuating part of the current. The point contacts
emit fluctuations $\delta i_j^{\alpha}$ and $\delta i_S$, directly
into the reservoir as well as into the dot. As a consequence the
electron and hole distribution functions aquires a fluctuating
part,\cite{vanlangen97,Blanter00} $f^{\alpha}(t)=\bar
f^{\alpha}+\delta f^{\alpha}(t)$. The total fluctuating current in
each contact is given by (suppressing the time notation)
\begin{eqnarray}
\Delta i_{j}^{\alpha}=\delta i_{j}^{\alpha}+ N\Gamma\delta
f^{\alpha},~\Delta i_S=\delta i_S+MR(\delta f^e-\delta
f^h).
\label{fluctdens}
\end{eqnarray}
The fluctuating parts of the distribution functions, $\delta f^e$ and
$\delta f^h$, are determined from the demand that the fluctuating
current, in the zero frequency limit considered, is conserved in each
dot,
\begin{eqnarray}
\Delta i_1^e+\Delta i_2^e+\Delta i_S=0,~\Delta i_1^h+\Delta
i_2^h-\Delta i_S=0.
\label{curr3}
\end{eqnarray}
Solving these two equations gives $\delta f^e$ and $\delta f^h$, which
can be substituted back into Eqs. (\ref{fluctdens}) to give the
fluctuating currents in the leads $\Delta i_j^{\alpha}$ in terms of
the fluctuating currents emitted by the point contacts $\delta
i_j^{\alpha}$ and $\delta i_S$. The charge current fluctuations in
contact $1$ and $2$ is then found by subtracting the electron and the
hole currents, i.e. $\Delta i_j=\Delta i_j^e-\Delta i_j^h$, giving
\begin{eqnarray}
\Delta i_1=\frac{(N\Gamma+2MR)[\delta i_1^e-\delta i_2^h]+N\Gamma(\delta i_1^h-\delta i_2^e+2\delta i_S)}{2(N\Gamma+MR)} \nonumber \\ 
\label{deltai12}
\end{eqnarray}
and $\Delta i_2=\Delta i_{1\rightarrow 2}$. For the second moment, the
point contacts can be considered \cite{cumulantcom} as independent
emitters of fluctuations, i.e. (here including $\delta i_S$)
\begin{eqnarray}
\langle \delta i_{j}^{\alpha}(E,t)\delta
i_{k}^{\beta}(E',t')\rangle=\frac{h}{e}\delta_{jk}\delta_{\alpha\beta}\delta(E-E')\delta(t-t')S_{j}^{\alpha}
\label{cumulants}
\end{eqnarray}
where the fluctuation power $S_{j}^{\alpha}(E)$ is determined by the
transparency of contact $j$ and the time averaged distribution
functions on each side of the contact, \cite{Buttiker92} as
\begin{eqnarray}
S_{j}^{\alpha}(E)&=&eN\Gamma[f^{\alpha}_{j}(1-f^{\alpha}_{j})+\bar f^{\alpha}(1-\bar f^{\alpha}) \nonumber \\
&+&(1-\Gamma)(\bar f^{\alpha}-f^{\alpha}_{j})^2], \nonumber \\ 
S_S(E)&=&eMR[\bar f^e(1-\bar f^e)+\bar f^h(1-\bar f^h) \nonumber \\
&+&(1-R)(\bar f^e-\bar f^h)^2]. 
\label{powerspectra}
\end{eqnarray}
Eqs. (\ref{currcorrexpr}) to (\ref{powerspectra}) form a complete set of
equations to calculate the current correlations.

\begin{figure}[h]        
\centerline{\psfig{figure=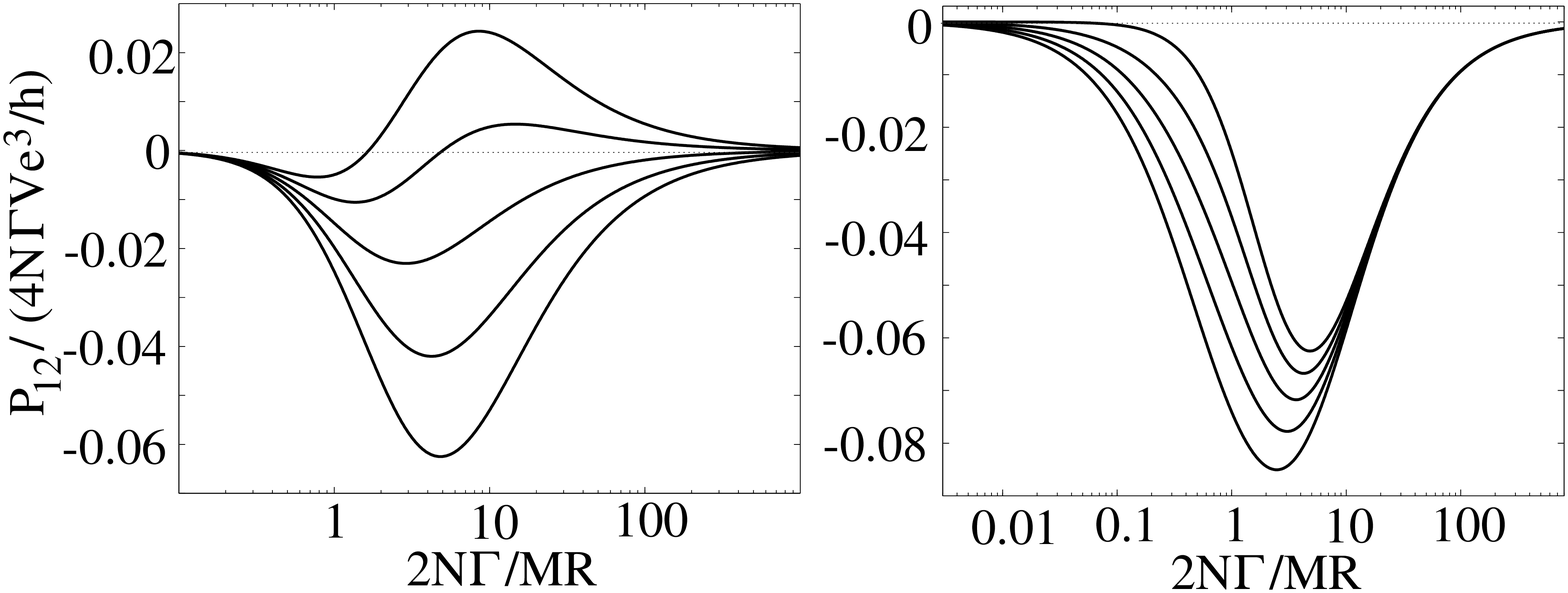,width=8.5cm}}
\caption{The cross correlation $P_{12}$ as a function of the ratio of
the conductances of the point contacts to the normal and
superconducting reservoirs, $2N\Gamma/MR$, for different barrier
transparencies. Left: $\Gamma=1$ and from bottom to top
$R=1,0.8,0.6,0.4$ and $0.2$. Right: $R=1$ and from top to bottom,
$\Gamma=1,0.8,0.6,0.4$ and $0.2$.}
\label{fig2}
\end{figure}

As an example, which also allows a comparison to
Ref. [\onlinecite{Samuelsson02}], we study the current cross
correlations $P_{12}$ for $V_1=V_2=V$ and $kT=0$. In this limit only
the energy interval $0<E<eV$ is of interest, where $f^e_1=f^e_2=1$ and
$f^h_1=f^h_2=0$. Correlating the current density fluctuations $\Delta
i_1$ and $\Delta i_2$ in Eq. (\ref{deltai12}) and using the
expressions for the power spectra in Eq. (\ref{powerspectra}), with
the distribution functions $\bar f^e$ and $\bar f^h$ taken from
Eq. (\ref{distfunc}), we obtain from Eq. (\ref{ehcurr}) and
(\ref{currcorrexpr}) for the the cross-correlator
\begin{eqnarray}
P_{12}&=&V\frac{e^3}{h}\frac{N^2M\Gamma^2R}{(N\Gamma+MR)^4}[2\Gamma^2N^2(1-2R) \nonumber \\ 
&-& MR(N\Gamma[2-\Gamma]+2MR[1-\Gamma])].
\label{result1}
\end{eqnarray}
In the limit without barriers, $\Gamma=1$ and $R=1$, as well as in the
limit $\Gamma=1$ and $N \gg M$, this expression coincides with the
random matrix theory result available in
Ref. [\onlinecite{Samuelsson02}]. From Eq. (\ref{result1}), we can
make the following observations. In the limit of dominating coupling
to the superconductor, $MR \gg N\Gamma$, the cross correlations are
manifestly negative (see Fig. \ref{fig2}). The positive correlations
predicted in this limit in Refs. [\onlinecite{Boerlin02,Samuelsson02}]
are thus due to the proximity effect.
In the opposite limit, dominating coupling to the normal reservoirs
$N\Gamma \gg MR$, the correlations are $P_{12}\propto R(1-2R)$,
positive for $R<1/2$.

In the general case, with different transparencies or widths of the
two contacts to the normal reservoirs, an asymmetric bias $V_1 \neq
V_2$ or finite temperatures, a detailed study shows that the
conditions $R<1/2$ and a dominating coupling of the dot to the normal
reservoirs are still necessary, but not always sufficient, for
positive cross correlations. The junction parameters of the example
studied above are the most favorable for obtaining positive
correlations.
 
We now show that the result in Eq. (\ref{result1}) can be obtained by
an ensemble averaged, quantum mechanical Green's function approach,
when supressing the proximity effect with a weak magnetic field in the
dot. We apply the circuit theory of
Refs. [\onlinecite{Nazarov,Boerlin02}], which is formulated to treat
the full counting statistics of the charge transfer. The formulation
of the problem is similar to Ref. [\onlinecite{Boerlin02}], so we keep
the description short. However, we treat arbitrary contact
transparencies and finite magnetic field in the dot.

\begin{figure}[h]        
\centerline{\psfig{figure=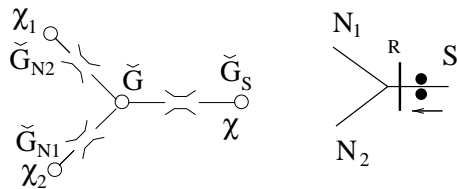,width=6.0cm}}
\caption{Left: The circuit theory representation of the junction with
the Greens functions of each node and the counting fields
shown. Right: The particle counting model. (see text)}
\label{fig3}
\end{figure}

A picture of the circuit is shown in Fig. \ref{fig3}. It consists of
four ``nodes'', the normal and superconducting reservoirs and the dot
itself, connected by ``resistances'', the point contacts.  Each node
is represented by a $4\times4$ matrix Greens functions (see
Fig. \ref{fig3}).
The Greens functions of the reservoir nodes are known, given by
$\check G_{Nj}=\mbox{exp}(i\chi_j\check \tau_K/2) \check
G_{N}\mbox{exp}(-i\chi_j\check \tau_K/2)$ and $\check
G_{S}=\mbox{exp}(i\chi\check \tau_K/2)\bar 1 \hat
\sigma_x\mbox{exp}(-i\chi\check \tau_K/2)$, where $\check
G_{N}=\bar\tau_z\hat\sigma_z+(\bar\tau_x+i\bar \tau_y)\hat1$ and
$\check \tau_K=\bar\tau_x\hat\sigma_z$. Here $\bar \tau (\hat \sigma)$
are Pauli matrices in Keldysh (Nambu) space and $\chi_1, \chi_2$ and
$\chi$ are counting fields, ``counting'' the number of electrons
entering and leaving the reservoirs. The unknown Green's function
$\check G(\chi_1,\chi_2,\chi)$ of the dot, normalized as $\check
G^2=1$, is determined from a matrix version of the Kirchoffs rule,
\begin{eqnarray}
\check I_{N1}+\check I_{N2}+\check I_{S}+\check I_{\Phi}=0
\label{curreq}
\end{eqnarray}
where the matrix currents $\check I_{N1}, \check I_{N2}$ and $\check
I_{S}$ are given by
\begin{eqnarray}
\check I_{N1(2)}&=&N\Gamma[\check G_{N1(2)},\check
G](4+\Gamma[\{\check G_{N1(2)},\check G\}-2])^{-1} \nonumber \\ \check
I_{S}&=&M\Gamma_S[\check G_{S},\check G](4+\Gamma_S[\{\check
G_{S},\check G\}-2])^{-1},
\label{matcurr}
\end{eqnarray}
where $[A,B] (\{A,B\})$ is the (anti)commutator. The last current
term, dominated by the presence of a magnetic field in the dot, is
\cite{Volkov} $I_{\Phi}=\gamma (h\Phi/e)^2 [\check\tau_0 \check G
\check\tau_0,\check G]$, where $\Phi$ is the magnetic flux in the dot,
$\check \tau_0=\bar1\hat\sigma_z$ and $\gamma$ is a constant of order
of $2N\Gamma,M\Gamma_S$, i.e. much larger than one. We are interested
in the case with suppressed proximity effect, which can be achieved by
applying a magnetic field corresponding to a flux $\Phi$ larger than a
flux quantum. This implies that the $I_{\Phi}$ term in
Eq. (\ref{curreq}) is much larger than the other terms, and
consequently that $[\check \tau_0\check G\check \tau_0,\check G]=0$ to
leading order in $1/\Phi^2$. The cross correlator $P_{12}$ is
\begin{eqnarray}
P_{12}=Ve^3/h\mbox{tr}\left.\left[\check \tau_K \partial \check I_{N1}/\partial
\chi_2\right]\right|_{\chi_1=\chi_2=\chi=0}.
\label{crosscorrcirc}
\end{eqnarray}
It is not possible to find an explicit expression for $\check
G(\chi_1,\chi_2,\chi)$ from Eq. (\ref{curreq}) and the additional
conditions $\check G^2=1$ and $[\check \tau_0\check G\check
\tau_0,\check G]=0$. However, to evaluate the correlator $P_{12}$,
only $\partial \check I_{N1}/\partial \chi_2$ is needed, and
consequently only the Green's functions expanded to first order in
$\chi_2$. We get $\check G=\check G^{(0)}+i(\chi_2/2)\check G^{(1)}$,
where $\check G^{(n)}=\partial^n \check
G/\partial\chi_2^n|_{\chi_1=\chi_2=\chi=0}$, and similar for the
others. From Eq. (\ref{curreq}), expanded to first order in $\chi_2$
as well, we then arrive at equations for $\check G^{(0)}$ and $\check
G^{(1)}$.

The physically relevant result for $\check G^{(0)}$ is
$\bar\tau_z\hat\sigma_z+h(\bar\tau_x+i\bar \tau_y)\hat1$, where
$h=N\Gamma/(N\Gamma+MR)$. Knowing $\check G^{(0)}$, and using that
from $\check G^2=1$ we obtain $\{\check G^{(0)},\check G^{(1)}\}=0$,
we then get $\check G^{(1)}=-h^2\bar \tau_z \hat \sigma_z+\tilde
h(\bar\tau_x+i\bar \tau_y)\hat1+h(\bar\tau_x-i\bar \tau_y)\hat1$,
where $\tilde
h=-h[(1-2h+2h^2)+\Gamma(h-1)^3+2h^3MR(2R-1)/(N\Gamma)]$. Inserting the
expressions for $\check G^{(0)}$ and $\check G^{(1)}$ into $\partial
I_{N1}/\partial \chi_2$, we find $P_{12}$ from
Eq. (\ref{crosscorrcirc}). This gives exactly the semiclassical result
in Eq. (\ref{result1}).

Finally, it was pointed out that in the limit of dominating coupling
to the normal reservoirs, $N\Gamma \gg MR$, the simple expression for
the cross correlations $P_{12} \propto R(1-2R)$ could be explained by
particle counting arguments. \cite{Samuelsson02} We now show that the
full probability distribution of transmitted charges in this limit can
be derived from the circuit theory. The total cumulant generating
function $F(\chi_1,\chi_2,\chi)$ is the sum of the functions for each
point contact, $F=F_{N1}+F_{N2}+F_S$, where
\begin{eqnarray}
&&F_{S}=eVM\Gamma_S ~\mbox{tr}~\mbox{ln}[4+\Gamma_S(\{\check G,\check
G_{S}\}-2)], \nonumber \\ 
&&F_{N1(2)}=eVN\Gamma
~\mbox{tr}~\mbox{ln}[4+\Gamma(\{\check G,\check G_{N1(2)}\}-2)].
\label{CGF}
\end{eqnarray}
We expand the Green's function of the dot, $\check G$, to first order
in $MR/N\Gamma$, (but no expansion in the counting fields). To zeroth
order in $MR/N\Gamma$, the superconducting reservoir is disconnected,
and we get directly $\check G^{(0)}=(\check G_{N1}+\check
G_{N2})/2$. Since normalization implies $\{\check G^{(0)},\check
G^{(1)}\}=0$, we note that the contribution to $F$ from the point
contacts connected to the normal reservoirs, $F_{N1}+F_{N2}\propto
\{\check G^{(0)},\check G^{(1)}\}$ disappears to first order in
$N\Gamma \gg MR$. The total $F(\chi_1,\chi_2,\chi)$ is then obtained
by inserting $\check G=\check G^{(0)}$ into Eq. (\ref{CGF}), giving
(for $V<0$)
\begin{eqnarray}
F=eVM~\mbox{ln}\left[1-R+\frac{R}{4}\left(e^{i(\chi_1-\chi)}+e^{i(\chi_2-\chi)}\right)^2\right].
\label{cumgenfunc}
\end{eqnarray}
In the limit $R\ll 1$, $F$ has the same form as in
Ref. [\onlinecite{Boerlin02}]. The corresponding probability
distribution $P(Q,q_1,q_2)$ of Eq. (\ref{cumgenfunc}), that $q_1(q_2)$
particles have left the dot through contach 1(2) when $Q$ pairs have
attempted to enter the dot, is (for even $q_1+q_2$) given by
\begin{eqnarray}
&&P(Q,q_1,q_2)=\frac{Q!(q_1+q_2)!}{q_1!q_2!(Q-[q_1+q_2]/2)!([q_1+q_2]/2)!}
\nonumber \\
&&\times\left(\frac{R}{4}\right)^{(q_1+q_2)/2}(1-R)^{Q-(q_1+q_2)/2}.
\label{supdist}
\end{eqnarray}
This is just the distribution function one gets from the model of
Ref. \onlinecite{Samuelsson02}. A filled stream of pairs are incoming
from the superconductor. Each {\it pair} has a probability $R$ to
enter the dot and then each {\it particle} in the pair has an
independent probability $1/2$ to exit through one of the contacts $1$
or $2$ (see Fig. \ref{fig3}).

The processes where both particles exit through the same contact
contributes $\propto -R^2$ to the cross correlations, while the
processes where the pair breaks and one particle exits throgh each
contact contributes $\propto R(1-R)$. Thus, for $R<1/2$, the pair
breaking noise dominates over the pair partition noise and the cross
correlations are positive. We emphasize that it is the fact that the
superconductor emits particles in pairs that makes positive
correlations possible, in a corresponding normal systems the electrons
tries to enter the dot one by one and the cross correlations are
manifestly negative.

In conclusion, we have presented a semiclassical theory of
current-current correlations in multiterminal chaotic
dot-superconducting junctions. It is found that the current cross
correlations are positive for finite backscattering in the
dot-superconducting contact and dominating coupling of the dot to the
normal reservoirs. We have also shown that this approach is equivalent
to an ensemble averaged Greens function approach with a suppressed
proximity effect

We acknowledge discussions with W. Belzig, Y. Nazarov, H. Schomerus
and E. Sukhorukhov. This work was supported by the Swiss National
Science Foundation and the program for Materials with Novel Electronic
Properties.


\begin{thebibliography}{999}
\bibitem{commentvolt} 
The approach can be extended to arbitrary
voltages and temperatures, by taking into account the energy dependence of the
Andreev refelction probability as well as single particle transport into the superconductors at $|E|>\Delta$ (the latter done via an additional contact).
\end{thebibliography}
\end{document}